\documentstyle[new_cite,times]{mn}

\title{RXTE observations of the Seyfert galaxy NGC 5506: 
evidence for reflection from disk and torus}
\author[]
  {G.~Lamer,\thanks{E-mail: gl@astro.soton.ac.uk} 
  P.Uttley, and I.M.~M$^{\rm c}$Hardy \\
  $^1$Department of Physics and Astronomy, The University,
      Southampton, SO17 1BJ}

\date{Accepted. Received}
\pagerange{\pageref{firstpage}--\pageref{lastpage}}
\pubyear{1999}

\pagerange{\pageref{firstpage}--\pageref{lastpage}}
\pubyear{1999}

\begin{document}

\input{psfig}

\maketitle

\label{firstpage}

\begin{abstract}
We report on observations of the narrow line Seyfert galaxy NGC 5506 with
the {\it Rossi X-ray timing explorer}. The observations cover a time
interval of $\sim$1000 days during which the source showed strong flux
and spectral variability. The spectrum clearly shows iron 
fluorescence emission at 6.4 keV and significant reflection features.
Both the equivalent width of the iron line and the relative strength
of the reflected continuum are higher during  low flux states.
The variability of the reflection features can be explained by 
the presence of two reflected components: one that responds rapidly to flux 
changes  of the primary continuum and a second, slowly variable, 
component originating from a distant reflector, e.g. a molecular torus.

\end{abstract}


\section{Introduction}

NGC~5506 is a bright nearby (z=0.006) narrow emission line X-ray galaxy. 
It has been classified variously as edge on early type spiral 
\cite{deV} and irregular (\ncite{Wilson76}, \ncite{Wilson85}).
Strong nuclear dust lanes  as well as starburst activity are present
in this galaxy. 
Although the optical spectrum of the nucleus is classified as Seyfert
2, NGC~5506 is not a 
typical member of this subclass.  All spectral components, including
the narrow lines, are strongly reddened by the disk of the host
galaxy. The X-ray source is absorbed by a column of 
$N_{\rm H}=3.6 \cdot 10^{22}$ \cite{Perola}. This column is moderate for a Seyfert 2
nucleus and might be largely due to absorption by interstellar matter in
the edge-on galaxy, rather than absorption by a nuclear torus. 
Despite this relatively low absorption
in  X-rays, no broad line region has been unambiguously detected 
in near infrared observations \cite{Veilleux}. \scite{Goodrich} conclude that 
the relative broadening of the NIR lines relative to the optical lines
is due to a contribution from the inner parts of the narrow line
region that is reddened by the dust lane crossing the bulge of NGC 5506.
 
The 2-30 keV X-ray spectra of Seyfert galaxies are dominated by two 
components: a power law component that is believed to originate from
the primary source of hard X-rays near the central black hole and
a reflected component produced by Compton scattering of the primary
radiation by neutral material (eg \ncite{Pounds}). 
The  principal spectral signatures 
of the reflected component are  a broad  hump peaking
at $\sim 30$ keV and iron $K_{\alpha}$ fluorescence emission at 6.4 keV.
The strength of the broad hump is more or less independent of the
relative abundances but the strength of the iron feature depends
directly on the iron abundance.

In many Seyfert X-ray spectra a broad, gravitationally redshifted iron 
fluorescence line is present, indicating that the reflection at 
least partly occurs in the accretion disk, (\ncite{Tanaka},
\ncite{Nandra97}, \ncite{Reynolds}).  
However, \scite{Ghisellini} and \scite{Krolik} showed that  the
molecular torus  required
by unified models of Seyfert galaxies is also likely  to modify 
the  X-ray spectra of both Seyfert 1 and  Seyfert 2 galaxies by
reprocessing the primary radiation. The molecular torus is
the supposed origin of narrow iron $K\alpha$ components found in the
{\it ASCA} spectra of several Seyfert galaxies (\ncite{Weaver97}, 
\ncite{Yaqoob}, \ncite{Weaver98}).
Any reflection component originating from the torus will respond only
slowly to variations of the primary X-ray source and therefore its 
{\it relative} contribution to the X-ray spectrum will be
anti-correlated  with the flux from   the variable X-ray source.
Due to this temporal behaviour reprocessing by a distant reflector can
be identified using long term monitoring X-ray observations.
The presence of an obviously distant reflector in the 
Seyfert 1 galaxy NGC 4051 was revealed when a
``bare'' reflection component became visible while the  primary X-ray
source virtually switched off for $\sim$ 150 days (\ncite{Guainazzi}, 
\ncite{Uttley}).

In this paper we present X-ray monitoring observations of NGC 5506
obtained with the {\it Rossi X-ray Timing Explorer (RXTE)} as well as a 
100 ksec {\it RXTE} observation spread over 20 days in June/July
1997. 
The reflection hump and
the 6.4 keV fluorescence line are clearly detectable in both sets of 
observations. 
Since both the equivalent width of the iron line and the relative reflected
fraction of continuum are higher during lower flux states, we 
conclude that two reflection components are present in NGC 5506. 

In Section \ref{obs} we describe the observations and the analysis of
the {\it RXTE} data.  The results of our spectral modelling are presentend
in Section \ref{spectra}. In Section \ref{specvar} we then describe the
variability of the spectral parameters and show that the observed
spectral variability can be described by a model involving both disk and
torus reflection.

\section{Observations and data analysis}
\label{obs}

\begin{figure*}
\par\centerline{\psfig{figure=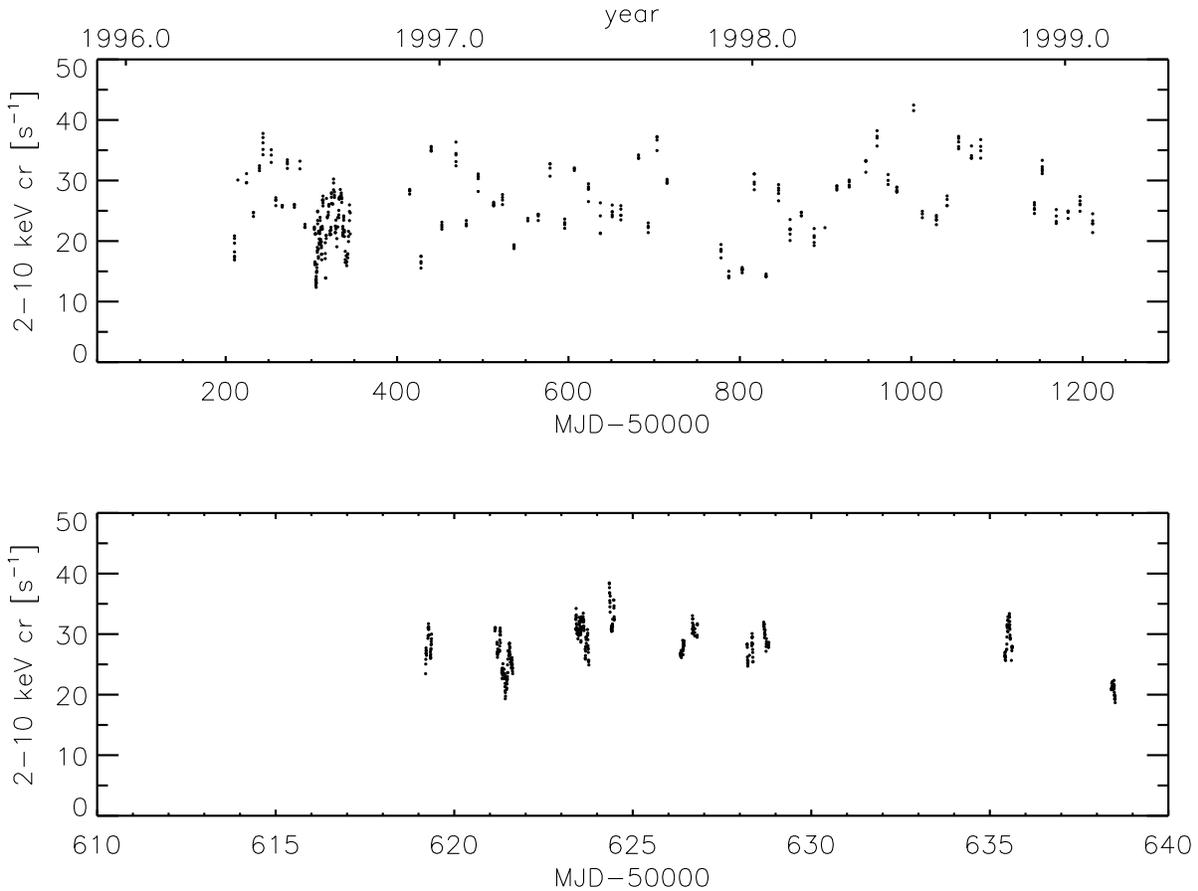,width=17truecm}}
\caption{\label{lc}
{\it RXTE} Background subtracted PCA 2-10 keV lightcurves during the long term 
monitoring campaign (upper panel) and the June 1997 observations (lower panel). Data from
PCUs 0-2 have been added, the binning is 256 sec. The errors in the
count rates are generally $< 0.5 {\rm s}^{-1}$, comparable to the
size of the plot symbols. Note that apparently simultaneous data
points in upper plot correspond to several time bins within one
observation, these data points give an indication on the short time
variability of the source.}
\end{figure*}

{\it RXTE} observed NGC 5506 with the Proportional Counter Array (PCA)
and  the High Energy X-ray Timing Experiment (HEXTE) instruments. 
The PCA \cite{Zhang} consists of 5 Xenon-filled Proportional Counter
Units (PCUs), sensitive to X-ray energies from 2-60 keV. The maximum
effective area of the PCA is 6500 cm$^2$. The HEXTE \cite{Rothschild}
covers the energy range 15-250 keV with a maximum effective area of 
1400 cm$^2$. 
We have observed NGC 5506 with the  {\it RXTE} since May 1996 as part of a
long term  monitoring campaign of 4 Seyfert galaxies.
We  used short ($\sim 1$ ksec) observations separated by a range of
time intervals. During 1996 the source was observed twice daily for 10 days,
daily for 32 days, and weekly for 13 weeks.
Since November 1996 the monitoring has continued with observations
every  two weeks.
The monitoring data presented here cover $\sim 1000$ days within 
the gain epoch 3 of the PCA instrument and amount to 77 ksec of good
exposure time.
We also observed NGC 5506 between 20 June 1997 and 9 July 1997 with
{\it RXTE} for 91 ksec. See Fig. \ref{lc}  for the
distribution of observations during both campaigns.

We have used {\sc ftools v4.2} for the reduction of the PCA and HEXTE
data. PCA ``good times'' have been selected from the Standard 2 mode data
sets using the following criteria: target elevation $> 10^{\circ}$,
pointing offset $<0.01^{\circ}$, time since SAA passage $> 30$ min,
standard threshold for electron contamination.
Since the PCUs 4 and 5 are often switched off, only the data from
the PCUs 0, 1, and 2 have been included.
Spectra and light curves were extracted from the top Xenon layer data.

\begin{figure}
\par\centerline{\psfig{figure=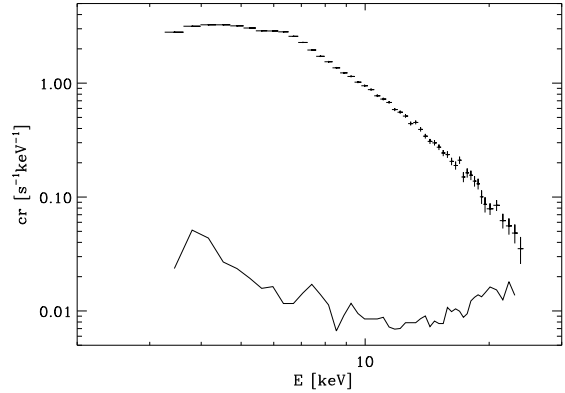,width=8truecm}}
\caption{\label{backest}
Comparison of a spectrum (data points) from the lowest states of NGC~5506 during the
long term monitoring (cr $< 21$ cts/sec, see Section \ref{specvar}) 
and the standard deviations in a set of background subtracted spectra 
of a blank field (solid line). The curve of standard deviations has been smoothed
for graphical representation.}
\end{figure}

We calculated the background in the PCA with the tool 
{\sc pcabackest v2.1} using the L7 model for faint sources, which 
is  suitable for determining the PCA background for energies $\leq 24$ keV.
In the case of NGC~5506, the PCA background dominates over the source 
spectrum at energies $>15$ keV. Therefore correct background 
subtraction is crucial for the determination  of the reflected 
fraction of the source spectrum  from the strength of the reflection hump at
energies $> 10$ keV.
In order to test the accuracy of the background model we analysed
100 PCA  blank field pointings  from observation P30801. 
We performed the data  reduction and background estimation for
the blank field pointings  in the same manner as for the
NGC~5506 science data   and produced a set of background subtracted 
spectra including PCUs 0-3. 
From this set of spectra  we derived the 
intrinsic standard deviation in each spectral channel by deconvolving the measured 
variance with the Poisson errors in the respective channel. 
The result represents the $1 \sigma$ systematic error in the spectra due to
inaccurate  background subtraction.
Figure \ref{backest} shows a comparison of a spectrum selected from the
faintest states of NGC~5506 (cr $<$ 21 cts/sec, see Section
\ref{specvar} ) and the $1 \sigma$ background standard deviations.
The systematic errors are generally smaller than the Poisson errors 
in the respective channels.
We added a typical background subtracted blank field spectrum to
the above mentioned faint state spectrum and fitted the {\sc pexrav} 
models described below to the original and the modified spectrum.
The resulting change in the spectral parameters is
$\Delta \Gamma=-0.018$ for the spectral index or $\Delta R = 0.14 $
for the reflected fraction.  
Note that the flux selected spectra discussed in Section
\ref{specvar}  are averaged from many 
pointings and therefore inaccurate background subtraction in
individual pointings should have only negligible influence on the 
results  derived from these spectra.

Spectral fitting was performed using {\sc xspec v10.0} using the latest
releases of the detector response matrices. 
All observations reported here have been carried out during the PCA gain
epoch 3. However, small temporal gain variations lead to changes 
in the detector response within this epoch. In order to account for 
these changes we used the tool {\sc pcarsp v2.37} to create 
individual response
matrices for the June 1997 observation and for each of the monitoring 
observations.

The inter-calibration of PCA and HEXTE regarding the normalization
of the spectrum is still uncertain and the relatively low count rate 
(2 cts/s per cluster)
for NGC~5506 in  the HEXTE detectors prevents an accurate measurement of
the spectral slope in the HEXTE band alone. 
Therefore the inclusion of the HEXTE spectra in the fitting did not
significantly improve the constraints on spectral parameters  and 
we only present the results of the PCA observations here.
Since  the spectral calibration of the softest PCA channels is somewhat
uncertain and the L7 background model is not recommended for energies above 24 keV, we 
only use the PCA data between 3 keV and 24 keV.
 
For the modelling of the spectrum we use the {\sc xspec} {\sc pexrav} code
\cite{MZ} which calculates a power law with exponential
cutoff at high energies reflected by a slab of neutral material. The
iron emission  feature is modeled by a Gaussian line.  
The spectrum in the PCA energy range is only marginally affected by 
photoelectric absorption and by the high energy cutoff of the
underlying power law. \scite{Perola} measured the X-ray spectrum of  
NGC 5506 in the broader energy range of the {\it BeppoSAX} instruments
and find that the absorbing column density of the host galaxy is 
$N_{\rm H}=3.6\cdot10^{22}{\rm cm^{-2}}$. Their lower limit for the 
high energy cutoff is $ E_{\rm max} = 300 {\rm keV}$.
In the following we adopt their values for $N_{\rm H}$ and $ E_{\rm max}$.

The reflected fraction $R$ is defined as $R=\Omega/(2\pi)$ where
$\Omega$ is the solid angle subtended by the reflector as seen from
the X-ray source. In the case of  an infinite flat
disk as reflector $R=1$ would be expected.   
The intensity of the reflected component also strongly depends on the
inclination of the reflecting slab. However, it is not possible to 
simultaneously
constrain both inclination  and the reflective fraction $R$. 
We therefore adopted an inclination of $40^\circ$ as suggested
by the profile of the iron K$\alpha$ fluorescence line  measured by 
{\it ASCA} \cite{Wang}.All elementary abundances were set to the solar values.

In order to investigate whether the flux variability of NGC~5506 
is accompanied by spectral changes, we have accumulated
PCA spectra during epochs of similar source count rates in the 2-10 keV
lightcurves as shown in Fig. \ref{lc} . For the 
June 1997 observations we set the  ranges
$< 24$ cts/sec, 24-27 cts/sec,  27-30 cts/sec, $>30$ cts/sec.
For the monitoring observations, where the spread in count rates is
larger, 6 ranges have been chosen:  
$<21$ cts/sec, 21-24 cts/sec, 24-27 cts/sec,  27-30 cts/sec, 30-33
cts/sec, and $> 33$ cts/sec.

Since the detector response matrices of the PCA change over the time span
of the monitoring observations, we have produced
response matrices for each flux selected spectrum. These matrices have
been constructed by averaging the matrices of the individual 
observations using weight factors according to the exposures 
contributing to the final spectra.    

Each of the spectra was the fitted with a {\sc pexrav + gauss} model
as discussed above, the results on spectral variability are given in
Section \ref{specvar}.

\section{X-ray spectra}
\label{spectra}

\begin{figure}
\par\centerline{\psfig{figure=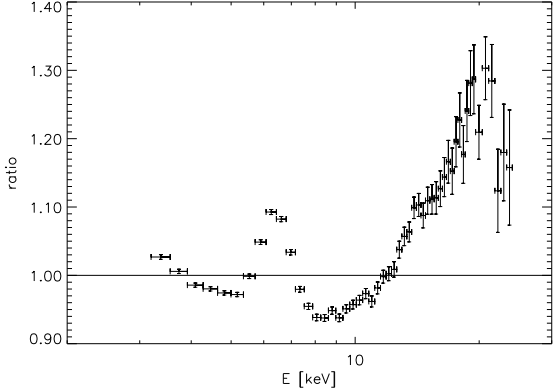,width=8truecm}}
\caption{\label{pwlfit}
Ratio of data to model when a simple power law spectrum is fitted
to the  June 1997 PCA spectrum of NGC 5506. Both the iron K$\alpha$
line and a reflection hump are obvious in the residuals.}
\end{figure}

A fit of the time averaged PCA spectrum from the June 1997 observation 
with a simple power law  model shows the presence of a strong 
Fe K$\alpha$ emission line 
and a significant reflection component at energies above $\sim 10$
keV (see Fig. \ref{pwlfit}).

The combination of the {\sc pexrav} model and a Gaussian iron line 
results in a satisfactory fit to the high signal to noise spectrum.
The iron line is slightly  resolved  with
$\sigma\sim0.30^{+.03}_{-.06}$ keV,
but  there is no evidence for any gravitational redshift. The best fit  
energy is  $6.404^{+.024}_{-.023}$ keV in the source frame
of NGC 5506.  This is consistent with higher resolution {\it ASCA}
spectroscopy by \scite{Wang},  who report $\sigma$=0.2 for a single 
Gaussian profile  
and no gravitational redshift of the iron $K\alpha$ line.     

The best fit value for the reflected fraction during the  June 1997 observation
is $R=1.2\pm0.08$. This is in good  agreement with the value measured 
by {\it BeppoSAX} in January 1997 ($R=1.05 \pm 0.58$, \ncite{Perola})
and is near the value $R=1$ that would be expected for the case of a 
simple geometry with a large disk reflecting the primary X-rays.

For the best fit {\sc PEXRAV} model the absorbed 2-10 keV flux  
is $1.02\cdot10^{-10}{\rm erg\; cm^{-2} s{-1}}$, the  corresponding 
X-ray luminosity is $1.25 \cdot 10^{42} {\rm erg\; s{-1}}$ 
($H_0=50 {\rm Mpc\; km^{-1} s}$, $q_0=0.5$).

\begin{figure}
\par\centerline{\psfig{figure=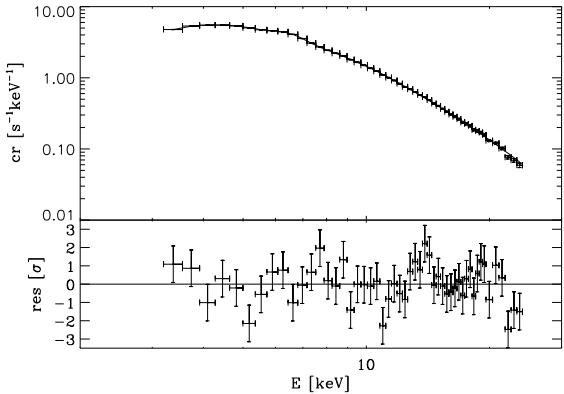,width=8truecm}}
\caption{\label{pexrav}
{\it RXTE} PCA spectrum fitted with a  {\sc pexrav+gauss} 
model. See Table \ref{specfit} for the model parameters.}
\end{figure}

\begin{table*}
\centering
\caption{\label{specfit} Fit parameters for time-averaged June 1997 spectrum}
\begin{tabular}{ @{}lcccccc@{} }
model     & $\Gamma$  & line EW & line E$_{\rm source}$ & line $\sigma$ & R$_{\rm refl}$ & $\chi^2$ (dof) \\
          &               &   [eV] & [keV]  &  [keV]            & &\\[10pt]

{\sc pwl}       &   2.00$\pm$.005 &  -     &  -     &   -               &  - & 3089 (48) \\
{\sc pwl+gauss} &   1.85$\pm$.007 &  248$\pm$15 & 6.265$\pm$.013& 0.41$\pm$.22&  - & 726 (45) \\
{\sc pexrav}    &   2.19$\pm$.008 &  -          &     -         &     & 1.766$\pm$.064& 1781 (48) \\
{\sc pexrav+gauss}& 2.13$^{+.01}_{-.01}$ &  149$^{+9}_{-6}$ & 6.404$^{+.024}_{-.023}$&      
0.302$^{+.032}_{-.058}$&1.20$^{+.08}_{-.08}$ &  52.7 (45) \\[10pt]
\end{tabular}
\end{table*}

\section{Spectral variability}
\label{specvar}

In our monitoring data from NGC 5506 we see variations in the 2-10 keV
count rate between 12 cts/sec and 42 cts/sec.  During the June 1997
observations the source varied between 19 cts/sec and 39 cts/sec.
We  analysed the  hardness-ratios between different  bands 
of the PCA energy range and  found
that the flux variations are accompanied by variations in the spectral
hardness: the spectrum appears to be softer during epochs of higher
flux. This spectral behaviour  has been observed  in a number of other
Seyfert galaxies before (e.g. \scite{McHardy}, \scite{Uttley98},
\scite{Chiang}.
In the framework of the reflection models there are two possibilities 
to explain an anti-correlation of source flux  and spectral hardness:

\begin{itemize}
\item[1.]  The intrinsic spectrum of the primary X-ray source might
soften during high flux states.

\item[2.] As the primary X-ray source weakens, the fraction of
reflected radiation might increase, e.g.  due to a time lag between
the primary and reflected components.    

\end{itemize}

In order to investigate these possibilities we fitted  count rate
selected spectra (as described in Section \ref{obs}) with  the {\sc
pexrav} \cite{MZ} model and a Gaussian emission line.
No significant variability of the source frame energy or the width
$\sigma$ of the iron  fluorescent line have been found. 
Therefore these parameters have been fixed to values compatible 
with any of the single fits, $E_{\rm source}=6.4$ keV and $\sigma=0.35$ keV. 
The results of the spectral fits are shown in Table \ref{fluxspec}.
The variations of intrinsic photon index, reflected fraction, and
iron line flux with the count rate are shown in Figure \ref{simplot}.
During the June 1997 observations the spectral index was  $\Gamma \sim
2.13$ and did not vary with flux. On the longer time scales of the
monitoring observations, however, an anti-correlation of flux and
hardness of the intrinsic power law is clearly visible. 
It is not clear whether the difference between the two data sets
regarding the variation of the spectral index is due to the different
time scales of the observations. An analysis of the interval of
more intense monitoring between MJD=50300 and MJD=50350 showed that
during this interval the source showed the same spectral index
variability as during the long term monitoring campaign as a whole.

\subsection{Reflection and iron fluorescence emission}

\begin{figure}
\par\centerline{\psfig{figure=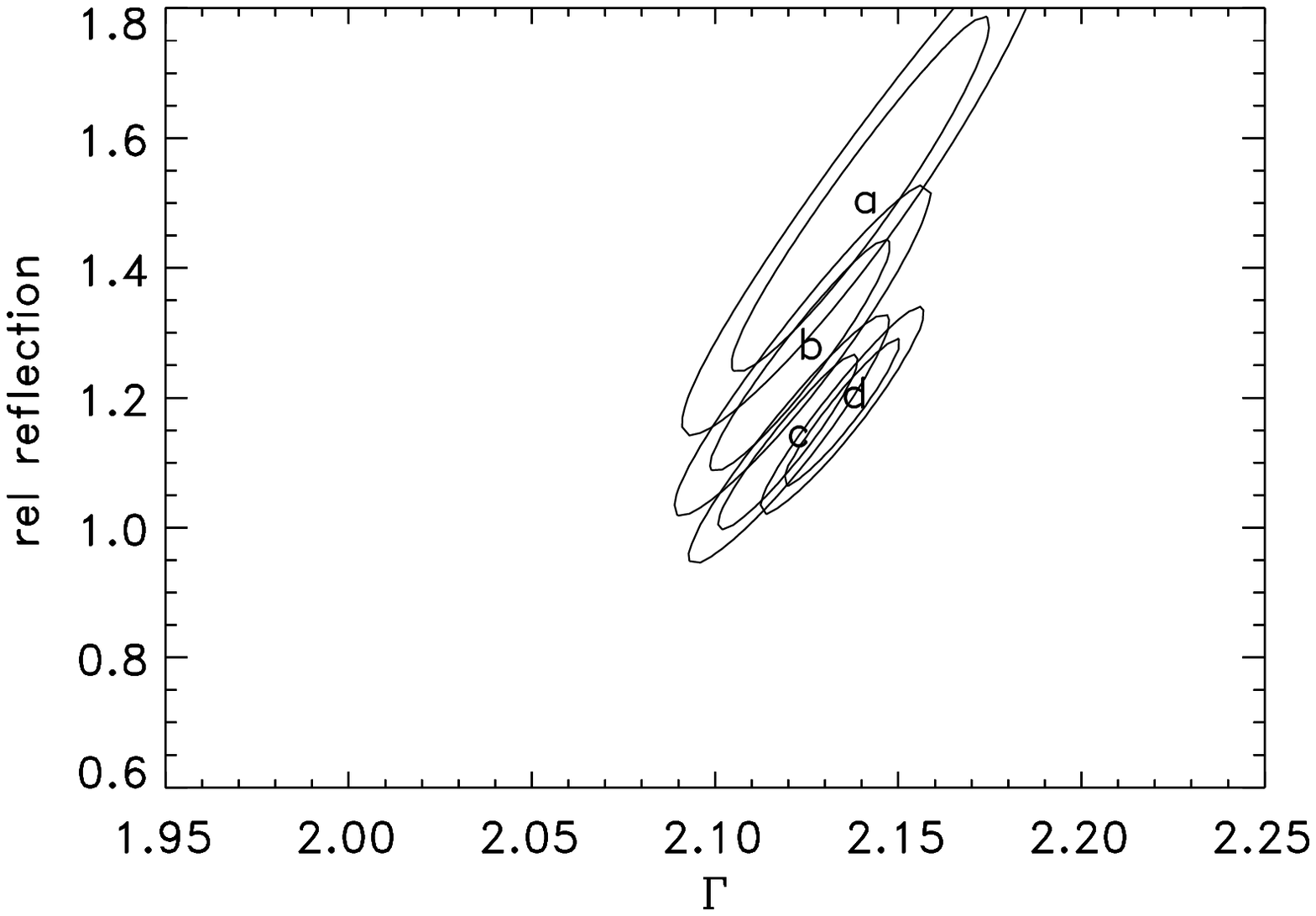,width=8truecm}}
\par\centerline{\psfig{figure=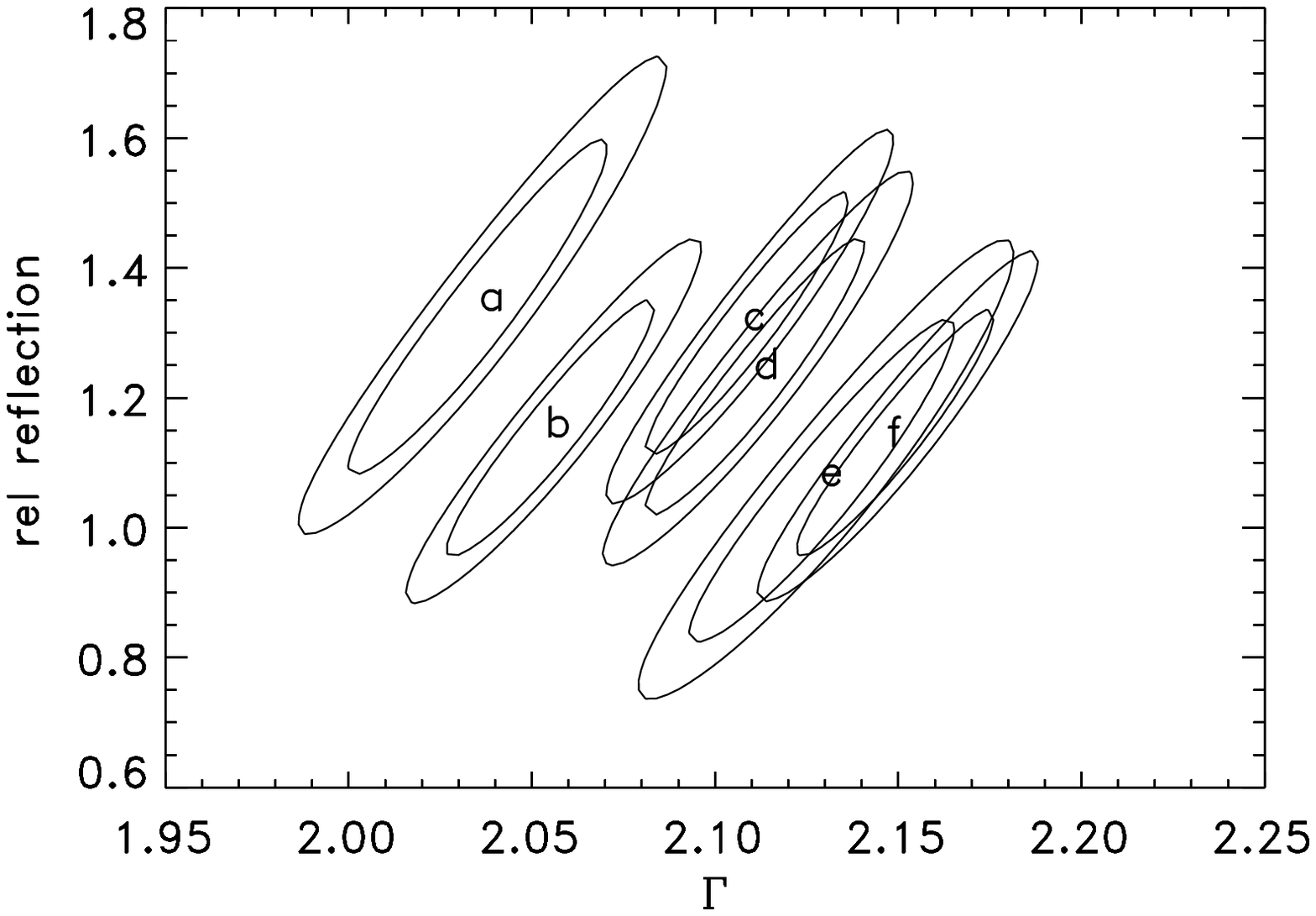,width=8truecm}}
\caption{ \label{contours}
68\% and 90\% confidence contours for the parameters $R$ (relative
reflection)  and $\Gamma$ (photon index) of the {\sc pexrav} model.
{\it Top:} Count rate selected spectra  from June 1997 observation, 
a) $<24 {\rm s^{-1}}$, b) $24 .. 27 {\rm s^{-1}}$, 
c)  $27 .. 30  {\rm s^{-1}}$, d)  $>30 {\rm s^{-1}}$.
{\it Bottom:} Count rate selected  spectra from monitoring
observations,
a) $<21 {\rm s^{-1}}$, b) $21 .. 24 {\rm s^{-1}}$, 
c)  $24 .. 37  {\rm s^{-1}}$, d)  $27 .. 24 {\rm s^{-1}}$, 
e) $30 .. 33 {\rm s^{-1}}$,
f) $>33 {\rm s^{-1}}$.
}
\end{figure}

The results on the reflected fraction for both the long term and 
the intense monitoring are consistent with each other and show
tentative evidence for an anti-correlation of  the reflected fraction
$R$ with source flux.
Figure \ref{contours} shows $68\%$ and $90\%$ contours for the 
parameters $\Gamma$ and $R$. On the shorter time scales  of the June
1997 observations the changes in reflected fraction $R$ contribute to
the variability in spectral hardness, while on longer time scales  
variations of the intrinsic power law index $\Gamma$ seem 
to be the main cause of the spectral variability. Variability of
the intrinsic spectral index was found to be the source of spectral
variability in the cases of NGC 5548 \cite{Chiang} and MCG$-6-30-15$
\cite{Lee}.
Although the best fit values of the  reflected fraction for the
different flux states in the long term monitoring are formally
consistent with each other, an anti-correlation with the source flux
is present. \scite{Zdziarski} report a {\it positive} correlation of  
$\Gamma$ and $R$ in samples of Seyfert galaxies and X-ray binaries and
claim that this correlation
also holds for the spectral variability of individual sources. 
Our results on NGC 5506 are in contrast to these findings, as  $\Gamma$
and $R$ are anti-correlated during the long term monitoring
observations. \scite{Zdziarski} explain the $R(\Gamma)$ correlation as
follows: The power law X-ray spectra arises in a hot Comptonizing
plasma surrounding a cool accretion disk. Cooling of plasma by soft
photons from the disk affects the spectral index. The correlation of
$R$ and $\Gamma$ is likely, as both the strength
of the reflection and the flux of cooling photons depend on the
properties of the disk. The absence of this correlation in the
spectral variability of NGC~5506 means that either the corresponding
properties of the accretion disk  are not variable in this source or 
the effect on the reflected fraction is compensated for by reflection
from the torus discussed in Section \ref{reflection}.

 Note that the anti-correlated variations in $R$ and $\Gamma$
presented  here cannot be
artifacts of the spectral fitting procedure, which may cause fake
correlations between two parameters in the direction of the 
major axes of the corresponding error ellipses. Spurious 
variations in  $R$ and $\Gamma$ due to their degeneracy 
in the {\sc pexrav} model would result in a positive correlation.
In fact the anti-correlation of $R$ and $\Gamma$ requires real
variations  in the hardness of the X-ray spectrum.

During the long term monitoring the  iron line flux does increase
with the continuum flux, but slower than proportionally as 
is obvious from the decreasing values of the equivalent width.
The spectra from June 1997 do not show significant variations of the
iron line flux, but the values are consistent with the results from
the long term monitoring. Therefore it is not clear whether the iron
line flux is actually constant on shorter time scales or whether the
variations are not detectable due to the narrower range of flux states 
during June 1997. 
The bottom panel of Figure \ref{simplot}  shows a significant increase
of the  iron line equivalent width at low source flux levels.
This is indicative of a constant component the iron $K_\alpha$ line.

\subsection{Disk and torus reflection}  
\label{reflection}
  
The variations of reflected fraction and iron line flux with the 
continuum flux indicate that two reflectors contribute to the 
reflection hump and the iron line. One reflector, most likely the
accretion disk,  is close to the central
source and responds rapidly to changes in the primary X-ray source.
A second reflection component, the tentative torus component, 
arises from matter more distant from the 
central source and hence is constant on the observed time scales.

Using the {\sc pexrav+gauss} model we  have simulated X-ray spectra for a
scenario  assuming the following components:

\begin{itemize}

\item[1.] Variable reflection that is proportional to the current 
intrinsic flux with the reflected fraction $R_{\rm disk}$.   

\item[2.] Constant reflection of an intrinsic  source 
whose luminosity corresponds to the average luminosity of 
the central X-ray source. The reflected fraction is $R_{\rm torus}$.

\item[3.] Each of the reflection  components are accompanied
by iron $K_{\alpha}$ fluorescence emission. The ratios of reflected
continuum flux and fluorescence flux   
are assumed to be the same in both components.
   
\end{itemize}

\begin{figure}
\par\centerline{\psfig{figure=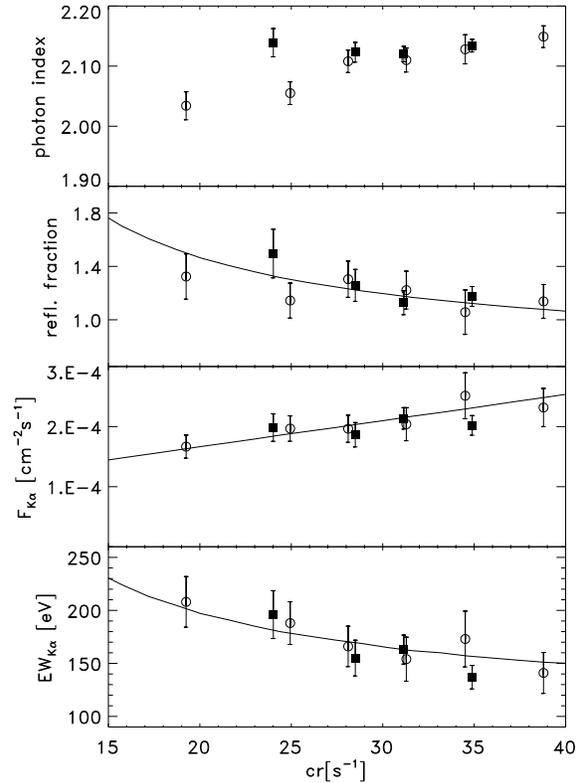,width=8truecm}}
\caption{\label{simplot}
Best fit  {\sc pexrav} model parameters as functions of the
count rate. From top to bottom:
 photon index $\Gamma$,
reflected fraction $R$, iron line flux, and iron line EW.
Filled squares: Results from June 1997 observations. 
Open circles: Monitoring observations. The lines 
indicate the values of reflected fraction and line
fluxes expected from a model with two reflectors: One close to the 
primary X-ray source and a more  distant reflector leading to a
non-variable reflected component (see Section \ref{reflection}.}
\end{figure}

\begin{figure}
\par\centerline{\psfig{figure=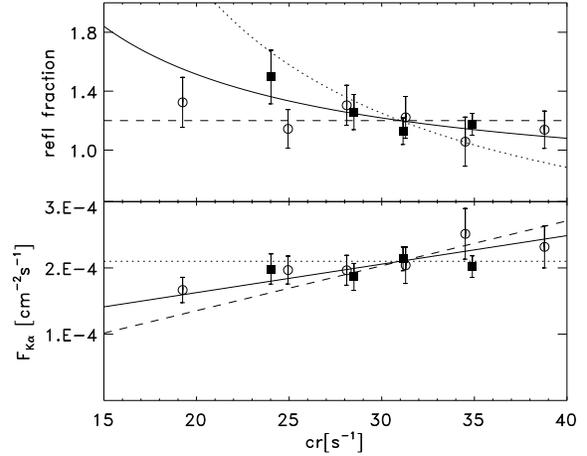,width=8truecm}}
\caption{\label{simcomp}
Best fit parameters $R$ and $F_{K\alpha}$  as in Fig. \ref{simplot}.
The solid lines indicate the ``two reflectors'' model described in 
Section \ref{reflection}. The dotted  lines show the expected
slopes for a single distant reflector, the dashed lines represent
a model with prompt reflection}       
\end{figure}

From a set of simulated spectra with various power law normalizations
the total count rates and line equivalents widths  have been
determined and the variations of the model parameters have been compared
with the observed values. The free parameters $R_{\rm disk}$, 
$R_{\rm torus}$,
and  fluorescence yield have been adjusted so that the simulations 
match the observations. The solid lines in Figure \ref{simplot} show 
the simulations with $R_{\rm disk}=0.7$ and  $R_{\rm torus}=0.5$.
The fluorescence yield corresponds to an equivalent width of 103 eV
for the disk reflection component alone.  
If the disk covers $\Omega=2\pi$ of the sky as seen from the source,
$R_{\rm disk}=1$ would be expected. 
For  disk reflection with inclination
$i=40^{\circ}$ and an intrinsic power law index of 
$\Gamma=2.1$ \scite{GF} calculated $EW_{K\alpha}=120$ eV.
Hence, both $R$ and  $EW_{K\alpha}$ from the disk component are 
of the right magnitude and the iron abundance is consistent with the
solar value.
According to our simulations the torus contributes an 
iron $K\alpha$ component of $EW=75$ eV when the source luminosity is 
at intermediate levels. This is in good agreement with the strengths
of the narrow Fe $K\alpha$ components found in {\it ASCA} spectra of
NGC~5506 (66 - 84 eV, \ncite{Wang})  and in a sample of other 
Seyfert 1.9 -- 2 galaxies ($<{\rm EW}>=$60 eV, \ncite{Weaver97}).  
This  and the satisfactory fit to the
observed spectral variability makes the disk+torus reflection scenario 
a likely model for NGC~5506. A single reflection component either from
the disk or from the torus is not consistent with the  observed spectral 
variability. Figure \ref{simcomp} shows the  reflected
fraction $R$ and iron line flux as functions of the count rate for 3
different models: the combined disk+torus reflection model as discussed
above (solid lines), a single torus reflector (dotted lines), and
prompt reflection from the accretion disk only (dashed lines). The
combined $\chi^2$ values when compared to the measured values of
$R$ and $F_{\rm K \alpha}$ are  $\chi^2/dof=12.2/17$ for the
`disk+torus' model,  $\chi^2/dof=56.6/18$ for the `torus only' model, and
$\chi^2/dof=29.3/18$ for the `disk only' model.

\begin{table*}
\centering
\caption{\label{fluxspec} Count rate selected spectra}
\begin{tabular}{ @{}lrcccccc@{} }
count rate & exposure &$\Gamma$  & line EW    & R$_{\rm refl}$ & $\chi^2$ (dof) \\
$[{\rm s}^{-1}]$&[s]  &           &   [eV]     &                 & \\[10pt]
           \multicolumn{6}{c}{June 1997}\\[10pt]
  $<24$ & 14592     & $2.139\pm 0.023$ & $196\pm 23$  & $ 1.496\pm 0.182$ & 34.5 (47)\\
 24-27  & 19760     & $2.123\pm 0.016$ & $155\pm 17$  & $ 1.258\pm 0.120$ & 32.6 (47)\\
 27-30  & 26352     & $2.120\pm 0.013$ & $163\pm 14$  & $ 1.128\pm 0.090$ & 34.0 (47)\\
 $> 30$ & 33744     & $2.134\pm 0.010$ & $137\pm 11$  & $ 1.175\pm 0.074$ & 42.0 (47)\\[10pt]

 \multicolumn{6}{c}{Long term monitoring}\\[10pt]
   $<21$ & 18112    & $2.040\pm 0.024$ & $207\pm 24$  & $ 1.358\pm 0.173$ & 42.8 (47)\\
   21-24 & 16688    & $2.059\pm 0.019$ & $194\pm 20$  & $ 1.172\pm 0.133$ & 39.3 (47) \\
   24-27 & 15616    & $2.113\pm 0.019$ & $172\pm 19$  & $ 1.334\pm 0.138$ & 35.7 (47) \\
   27-30 & 11168    & $2.114\pm 0.020$ & $154\pm 21$  & $ 1.247\pm 0.144$ & 29.9 (47) \\
   30-33 &  6064    & $2.132\pm 0.024$ & $165\pm 26$  & $ 1.080\pm 0.168$ & 23.0 (47)\\
   $>33$ &  9488    & $2.152\pm 0.018$ & $145\pm 19$  & $ 1.158\pm 0.128$ & 43.6 (47)\\
\end{tabular}
\end{table*}

\section{Conclusion}

We show that the variability of  the reflected continuum and 
the Fe $K\alpha$ fluorescence emission in the Seyfert 2 galaxy 
NGC~5506 are well described by a model that assumes reprocessing
by both an accretion disk and a distant medium, which might be the
molecular torus. 
In contrast to the monitoring observations of  NGC~4051 \cite{Uttley},
no long lasting states of low X-ray flux have been observed for 
NGC~5506. Therefore the distance of the reflecting matter from the
central X-ray source is not well constrained in this case.
However, the time delay of $>150$ days found for a similar constant reflection
component in NGC~4051 \cite{Uttley} supports the torus hypothesis.  

The relative contributions of the disk and the torus 
are approximately equal during the lowest flux states we have observed.
When the  X-ray luminosity is intermediate to high, reprocessing 
in the accretion disk dominates the reflected continuum and the
iron line.
The strength of the reflection features from the accretion disk alone
is broadly consistent with an inclination angle $i=40^{\circ}$ as 
inferred from {\it ASCA} measurements \cite{Wang} of the Fe
$K\alpha$ profile.  At this inclination the observed  X-ray absorption
($N_{\rm H}=3.6\cdot10^{22} {\rm cm^{-2}}$, \ncite{Perola}), which is
within the range for Seyfert 2 galaxies , but towards the lower end \cite{Turner} 
might be well below the total column density of the torus. 
The torus contribution to the  Fe $K\alpha$  fluorescence as inferred
from our variability study is in very good agreement with  
{\it ASCA} spectroscopy of the line profile \cite{Wang}. 
 
We conclude that long term monitoring of AGN is an 
excellent method to study the geometry of their inner regions as 
it has the potential to actually measure the linear scales of the
components contributing to reprocessing.

\label{lastpage}

\end{document}